\documentclass[11pt,twoside]{article}

\usepackage{asp2004}
\usepackage{epsf}
\usepackage{psfig}
\usepackage{lscape}
\usepackage{graphicx}
\usepackage{natbib}

\begin{document}
\title{Plausibility of the planet-engulfing scenario for V838 Mon from
the current knowledge on extrasolar planets}   
\author{S. Desidera}   
\affil{INAF-Osservatorio Astronomico di Padova}    

\begin{abstract} 

One of the hypothesis to explain the outburst of V838 Mon is the engulfment
of planets. More than 170 extrasolar planets were discovered in the past
years. Their properties and the characteristics of their host stars can be
used to evaluate of the plausibility of this hypothesis.
However, the large mass and young age of V838 Mon make the object rather 
different from
the typical targets of current planet searches.
Therefore, the expectations of planet engulfing events derived from 
them are not directly applicable to the case of V838 Mon.
Some of the properties of V838 Mon (as the probable large stellar mass
and sub-solar metallicity) make somewhat unlikely the presence 
of planets around it,
but the uncertainty on the mass and age of the progenitor and the lack
of knowledge of planet formation around massive star do not allow
a firm conclusion.

\end{abstract}

\section{Introduction}   

One of the hypothesis to explain the spectacular outburst of V838 Mon is 
the engulfment of planets (Retter \& Marom 2003; Retter et al.~2006).
More than 170 extrasolar planets were discovered in the past years
(Butler et al.~2006).
Their properties and the characteristics of their host stars may help
the evaluation of the plausibility of this hypothesis.
This approach is independent and complementary to the physical modeling of 
planet swallowing by different types of stars, aimed to study whether 
the unusual evolution of V838 Mon and related objects can be
explained by such an event.

We recall here some properties of V838 Mon that can be relevant
for a study of the possibility of having planets around it.

\begin{itemize}
\item
V838 Mon is a massive and young star. The details remain to be 
established because of the uncertainties on the photometry of the 
progenitor, on the contribution of the B3V companion to the 
integrated light of the system, and on reddening and distance.
Munari et al.~(2005)
favour a high mass of $\ge 30~M_{\odot}$ for the progenitor
and an age of a few Myr, on the basis of the location
of the progenitor blueward of the main sequence (see also 
R.~Hirschi, this conference) 
while Tylenda et al.~(2005)
propose two scenarios
depending on the adopted photometry of the progenitor and reddening and 
distance scales: a  $9~M_{\odot}$ star $<20$ Myr old or a $5~M_{\odot}$ 
in pre main-sequence phase (age 0.3 Myr).
It appears that there are no chance for the star to have been an AGB star 
in the past.

\item
V838 Mon is a binary star, as it has a B3V companion 
(Desidera \& Munari 2002).
This star remained apparently unaffected by the outburst, implying a
separation larger than a few AU.

\item
V838 Mon lies at a galactocentric distance of about 15 kpc.
At such galactic location a low metallicity for the young galactic
population is expected
([Fe/H] about -0.5; Maciel et al.~2005) 
and indeed suggested by the analysis of
Kipper et al.~(2004) and Kaminsky \& Pavlenko (2005).

\end{itemize}

\section{Properties of giant planets}   

The on-going searches for extrasolar planets have deeply changed our knowledge
of planetary systems, based until a few years ago only on the Solar System.
Now we know that giant planets can be found even at very small separations
from the parent stars and that planets with period longer than 10 days
(the limit of tidal circularization of the orbits) can have eccentricities
much larger than those of Solar System planets (Butler et al.~2006).
The mass function rises steeply toward lower masses, and there is likely
overlap in mass between objects formed in a circumstellar disk (planets) 
and objects formed via cloud fragmentation (brown dwarfs).
Multi-planet systems are rather common, showing a large variety of 
configurations.
Two systems (55 Cnc and GL 876) have two giant planets and one Super-Earth or 
Neptune-mass planet closely packed within 0.25 AU from the star 
(Mc Arthur et al.~2004; Rivera et al.~2005)
with a 
further planet in external orbit in the case of 55 Cnc, somewhat 
rensembling the planetary 
system architecture with three close giant planets postulated 
by Retter \& Marom (2003)
for V838 Mon. However, 
the uniqueness of these cases among more than 3000 stars under
high-precision radial velocity monitoring indicates an intrinsic
rarity of such a configuration.

\section{Properties of the host stars}   

The most relevant property of planet host is the high metallicity:
the frequency of planets is a strong function of the metallicity
of the parent star (Fischer \& Valenti 2005)
As an example, at [Fe/H]=-0.2 the frequency of giant planets is 
more than 6 times smaller than   that at [Fe/H]=+0.2.
Statistic concerning  metal poor stars is not sufficient to
determine if below [Fe/H]=-0.4 the planet frequency
continues to decrease or remains nearly constant
at lower metal content (Santos et al.~2004).
The lowest metallicity planet hosts have [Fe/H]$\sim -0.6$, if one
exlcudes the planet candidate around a pulsar in the globular cluster M4
([Fe/H]$\sim -1.2$), that may have been formed in different way
to planet found around solar-type stars (Sigurdsson et al.~2003).

Planets are found in binary systems with a large range of separations
(Eggenberger et al.~2004; Desidera \& Barbieri 2006),
indicating that planets can exists even in presence of fairly
strong dynamical interactions.

The frequency of planets as a function of stellar mass is only
poorly constrained from available data.
Infact, most of the planet search surveys are 
focused on solar-type stars (late F and G-K dwarfs).
A few surveys are considering M dwarfs, with preliminary indications
for a lower frequency of giant planets and a similar frequency
of Neptune-mass planets with respect to solar-type stars
(e.g. Bonfils et al.~2005).

\section{Planets and disks around intermediate and high mass stars}   

Search for planets around main-sequence stars more massive 
than 1.5 solar masses
is limited by the shallowness of absorption lines of early type
stars. Some planet search programs are focusing on their evolved
counterparts, i.e. giants, and a few planets have been discovered around 
K and G giants (Hatzes et al.~2006 and references therein).
The most massive planet host is probably HD 13189 ($3.5~M_{\odot}$);
however
the nature of the companion is somewhat uncertain as it has a mass at
the boundary between planets and brown dwarfs (Schuler et al.~2005).
More in general, the typically fairly large mass of planets 
orbiting stars more massive than 1.5 solar masses
might suggests that planet formation process scales with the primary mass,
but further studies are required to exclude the occurrence of
selection effects.
In any case, 
the mass range corresponding to the probable mass of the
progenitor of V838 Mon remains
basically unexplored by current planet search projects.

The metallicity of giant stars with planets is not anomalously high, 
possibly indicating that the planet-metallicity connection
observed for solar-type stars does not apply to intermediate mass stars.
This might be explained by intrinsic differences of planet 
formation processes at larger stellar masses (e.g. if  metal-poor disk
around massive stars reach adequate dust content to allow giant planet
formation while those round solar-type star do not).
However, selection effects in the radial velocity surveys of giant stars
might be at work (e.g. lower average metallicity of giants
included in radial velocity surveys, larger radial
velocity jitter and then lower planet detection capabilities for colder 
giants, color cuts in sample selection)
and should be carefully investigated.

Indirect evidence for the occurrence of planet formation around intermediate 
mass stars can be drawn from the presence of debris disks around a significant 
fraction of A type stars (Greaves \& Wyatt 2003).
Some late B type stars also posses debris-disks.
This indicates that planet formation proceeded
at least until the formation of planetesimals around these stars
(mass $\sim 1.6-3~M_{\odot}$).

Circumstellar disks around young stellar objects as massive 
as $\sim 10-20~M_{\odot}$ were observed, while
observational evidences of the presence of 
circumstellar disks
around stars more massive than $\sim20~M_{\odot}$ remain
elusive (Cesaroni et al.~2006).
In most cases, the observed
disks around $\ge 10-20~M_{\odot}$ stars are likely to be unstable 
and with short lifetimes (Cesaroni et al.~2006); possible long-lived
structures were recently reported around two B[e] hypergiants in the Large
Magellanic Cloud (Kastner et al.~2006). 
Formation of planets through the standard core-accretion
process in the circumstellar environment of massive stars
might be hampered by photoevaporation of dust grains.

\section{Planets engulfing} 

The period distribution and frequency of extrasolar planets implies
that a significant fraction of solar-type stars (about 3-5\%)
should engulf one or more giant planets during the RGB or AGB phases
(see e.g. Livio \& Soker 2002).

However, as noted in Sect. 1, the progenitor of V838 Mon was a rather warm
star,
likely with effective temperature higher than main sequence stars
of similar luminosity. The hypothesis of planet engulfing caused 
by an increase of stellar
radius driven by stellar evolution appears then unlikely.
Furthermore, the high temperature of the central star  
represents a severe challenge for the presence of a
Jupiter-mass hydrogen-rich planet orbiting
close to the star.
In fact, the UV flux of V838Mon progenitor was larger by several orders
of magnitude than that 
of a solar-like star, probably resulting in evaporation 
of gaseous planets at small separation (Lammer et al.~2003).

An additional channel to planet engulfing is represented by the 
dynamical interactions within a planetary system, that most likely
result in the ejection of planets from the system but sometimes cause
infall of planets into the central star (Marzari \& Wiedenschilling 2002).
A phase of dynamical instability of a planetary system most likely
arises at young ages, during the formation of the planetary system
(e.g. as the result of the runaway gas accretion on the icy cores,
if planets form according to the core-accretion scenario).
The timescales for the formation of giant planets 
in the core-accretion model are of a few Myr (Alibert et al.~2005),
comparable with the probable age of V838 Mon (Sect.~1).
In the case of V838 Mon, the presence of the massive B3V companion and
the interactions within the star cluster of which the star
is probably a member (Afsar \& Bond, this volume)
can also be a source of dynamical instability.

A planet engulfing event caused by dynamical instability appears more
suited to explain the high-energy outburst of V838 Mon, as
the impact velocity of the infalling planets should be larger
than in the case of planet engulfing caused by the increase of
stellar radius driven by stellar evolution, and
the planet could have been formed sufficiently far away to avoid
evaporation.

\section{Conclusions}   

The large mass and young age of V838 Mon make the object very different from
the typical targets of current planet searches.
Therefore, the expectations of planet engulfing events derived from 
them are not directly applicable to the case of V838 Mon. 
However, some indication for  the evaluation of the planet-engulfing
hypothesis  to explain V838 outburst can be derived.

\begin{itemize}
\item
An initial mass larger than $10-20~M_{\odot}$ 
would be critical for planet formation.
Planets can be found around such a massive star likely only if they form in 
a different way around
massive stars with respect to solar type stars (e.g. by disk instability).
\item
The young age of the system (a few Myr) is similar to the timescales
of planet formation around solar-type stars and then is 
compatible with the presence of planets.
\item
Dynamical instability of a young planetary system might cause
infall of a planet on the cental star with large impact velocity.
Planet engulfing caused by an increase of stellar radius appears
less viable because of the evolutionary status of the progenitor and 
the probable planet evaporation for a planet in stable orbit
close to the star
\item
The low metallicity of V838 Mon  works against the presence of
planets around it,  without being conclusive
\item
The binarity of V838 Mon does not exclude the presence of planets
\item
In any case the engulfment of three giant planets in two months during the 
outbust (Retter \& Marom 2003)
seems to 
require a very special architecture
of the planetary system, which appears very unlikely for mature planetary
systems and quite unlikely also for a scenario of  dynamical instability
during the early phases of a planetary system.
If lower mass planets or a single engulfment event might be responsible of
the outburst, tha plausibility of the scenario would become larger.
\end{itemize}

\acknowledgements 

I warmly thank Ulisse Munari for calling my attention on this
very special object and stimulating discussions on its nature.





\end{document}